\documentclass[fleqn,twoside]{article}
\usepackage{espcrc2}

\usepackage{graphicx}
\usepackage{epsf}
\usepackage{epsfig}
\usepackage{feynarts}
\usepackage[figuresright]{rotating}

\newcommand{\AmS}{{\protect\the\textfont2
  A\kern-.1667em\lower.5ex\hbox{M}\kern-.125emS}}

\title{Properties of Feynman graph polynomials}

\author{Christian Bogner\address[MCSD]{Institut f\"ur Theoretische Teilchenphysik und Kosmologie, 
        RWTH Aachen, \\ 
        D - 52056 Aachen, Germany}   \thanks{Talk given at the International Workshop 'Loops and Legs in Quantum Field Theory' (April 25-30, 2010, W\"orlitz, Germany).}}%

\begin{document}

\begin{abstract}
In this talk I discuss properties of the two Symanzik polynomials which characterise the integrand of an arbitrary multi-loop integral in its Feynman parametric form. Based on the construction from spanning forests and Laplacian matrices, Dodgson's relation is applied to derive factorisation identities involving both polynomials. An application of Whitney's 2-isomorphism theorem on matroids is discussed.
\vspace{1pc}
\end{abstract}

\maketitle

\section{INTRODUCTION}

The calculation of higher-order corrections in perturbative quantum field theory relies to some extent on a variety of techniques for the evaluation of Feynman integrals. Many of these techniques make use of Feynman parameters (see e.g.\ \cite{Smi}). In the Feynman parametric form, the integrand of a Feynman integral can always be expressed in terms of the two Symanzik polynomials  \cite{Itz,Nak,Todo}, called $\mathcal{U}$ and $\mathcal{F}$ in this talk. \\
Applying concepts of graph theory and matroid theory we discuss properties of these polynomials. Certain properties are already used explicitely in loop calculations while others are less regarded so far or previously unknown. Apart from the important role of the two polynomials in practical calculations they recently receive a lot of attention in a branch of the mathematical literature where Feynman integrals are considered as periods in the sense of algebraic geometry (e.g. \cite{BEK,BlKr,Pat,Mar,Sch,BeBr,BoWe1}). \\
Both polynomials are graph polynomials in the sense that they can be constructed directly from the Feynman graph. The first Symanzik polynomial  $\mathcal{U}$ is generated by the spanning trees of the Feynman graph and it can furthermore be derived as the determinant of a minor of a so-called Laplacian matrix, associated to the graph. Therefore a non-trivial identity of Dodgson on determinants is known to yield relations between first Symanzik polynomials belonging to different graphs. Using the all-minors matrix-tree theorem we derive the second Symanzik polynomial from a Laplacian matrix, where the external momenta are incorporated. Applying Dodgson's equation to the latter matrix we obtain an identity between first and second Symanzik polynomials of different graphs. \\
In the last section of our talk we ask for the conditions which two graphs need to fulfill to have the same first Symanzik polynomial. From the correspondence between spanning trees and bases of matroids we see that these conditions are given by a theorem of Whitney on cycle matroids.\\

\section{FEYNMAN PARAMETERS}

To an $l$-loop Feynman graph $G$ with $n$ internal edges  $e_1,\,...,\,e_n$ we associate the generic $D$-dimensional Feynman integral
$$I_G=(\mu^2)^{\nu -lD/2}\int \prod_{r=1}^l\frac{d^Dk_r}{i\pi^{D/2}}\prod_{j=1}^n \frac{1}{(-q^2_j+m^2_j)^{\nu_j}}. $$
To each edge $e_j$ we associate a mass $m_j$ (possibly zero) and a momentum $q_j$. Each $q_j$ is a linear combination of loop momenta $k_r$ and external momenta. The Feynman propagators are allowed to appear to arbitrary integer powers $\nu_j$ whose sum is denoted by $\nu=\sum_{i=1}^n \nu_i$. The arbitrary mass dimension $\mu$ is omitted in subsequent equations.\\
To each edge $e_i$ we furthermore assign a Feynman parameter $x_i$. Applying the well-known Feynman parameter technique (or 'trick') to each propagator and evaluating the integrals over the loop momenta, one arrives at the Feynman parameter integral \cite{Itz}
\begin{eqnarray*}
I_G & = & \frac{\Gamma(\nu-lD/2)}{\prod_{j=1}^n \Gamma (\nu_j)} \int_{x_j \geq 0}\left(\prod_{j=1}^n dx_j x_j^{\nu_j-1}\right)\\
& & \times \delta (1-\sum_{i=1}^nx_i)\frac{\mathcal{U}^{\nu-(l+1)D/2}}{\mathcal{F}^{\nu-lD/2}}. 
\end{eqnarray*}
The functions $\mathcal{U}$ and $\mathcal{F}$ are the objects of interest in this talk. They are polynomials in the Feynman parameters and $\mathcal{F}$ is furthermore a function of the kinematical invariants and squared masses of the Feynman graph.

\section{SPANNING FORESTS}

The polynomials $\mathcal{U}$ and $\mathcal{F}$ can be constructed directly from the Feynman graph. To this end one considers certain sub-graphs called spanning trees and spanning 2-forests. A forest is a graph without any loops (or cycles) and we speak of an $n$-forest if $n$ is the number of its connected components. A connected forest is called a tree. A spanning forest of a graph $G$ is a sub-graph of $G$ which is a forest and which contains the same vertices as $G$. If $l$ is the loop-number of $G$ then each spanning tree can be obtained by removing $l$ edges of $G$. Removing one more edge gives a spanning 2-forest in each case.\\
Let us denote the set of spanning trees of $G$ by $\mathcal{T}_1$ and the set of spanning 2-forests by $\mathcal{T}_2$. The Symanzik polynomials are obtained as follows \cite{Itz}: $$\mathcal{U}=\sum _{T\in \mathcal{T}_1} \prod _{e_i \notin T} x_i, $$
$$\mathcal{F}_0=\sum _{(T_1,\,T_2)\in \mathcal{T}_2} \left(\prod _{e_i \notin (T_1,\,T_2)} x_i \right)\left(-s_{(T_1,\,T_2)}\right) , $$
$$\mathcal{F}=\mathcal{F}_0+\mathcal{U}\sum_{i=1}^nx_im_i^2, $$
with 
$$s_{(T_1,\,T_2)}=\left(\sum _{e_j\notin (T_1,\,T_2)}q_j \right)^2.$$
Applying momentum conservation we consider the kinematical invariants $s_{(T_1,\,T_2)}$ as functions just of the external momenta. The polynomial $\mathcal{F}$ is expressed in terms of $\mathcal{U}$, $\mathcal{F}_0$ and a simple term involving the particle masses. Let us therefore restrict our attention to the polynomials $\mathcal{U}$ and $\mathcal{F}_0$. \\
$\mathcal{U}$ is a function of only the Feynman parameters while  $\mathcal{F}_0$ furthermore depends on the external momenta. We mention two further properties which can be directly obtained from the above equations:
\begin{itemize}
\item $\mathcal{U}$ and $\mathcal{F}_0$ are linear in each of the Feynman parameters (while $\mathcal{F}$ may contain parameters to the power of two). 
\item  $\mathcal{U}$ and $\mathcal{F}_0$ are homogeneous in the Feynman parameters of degree $l$ and $l+1$ respectively.
\end{itemize}
The above polynomials are generated by the edges which one removes in order to obtain the spanning trees and 2-forests respectively. Alternatively we can discuss the polynomials
$$U=\sum _{T\in \mathcal{T}_1} \prod _{e_i \in T} x_i, $$
$$F_0=\sum _{(T_1,\,T_2)\in \mathcal{T}_2} \left(\prod _{e_i \in (T_1,\,T_2)} x_i \right)\left(-s_{(T_1,\,T_2)}\right).$$ Note that they are generated by the edges belonging to the trees and 2-forests respectively. From the latter polynomials we obtain the previous ones by the simple transformations
$$\mathcal{U}(x_1,\,...,\,x_n)=x_1...x_nU(x_1^{-1},\,...,\,x_n^{-1}),$$
$$\mathcal{F}_0(x_1,\,...,\,x_n)=x_1...x_nF_0(x_1^{-1},\,...,\,x_n^{-1}).$$

\section{LAPLACIAN MATRICES}

Symanzik polynomials can be derived from matrices associated to graphs in various ways \cite{Itz,Nak,Kra}. The use of so-called Laplacian matrices is convenient for a direct application of theorems of the matrix-tree type. A Laplacian matrix $L$ of a graph $G$ with $r$ vertices is an $r\times r$ matrix. The diagonal entries are given by $L_{ii}=\sum x_k$, where the sum runs through the edges attached to the vertex $v_i$ and not being self-loops (or tadpoles), i.e.\ not attached to the same vertex at both of its ends. The off-diagonal entries of $L$ are $L_{ij}=-\sum x_k$, $i\neq j$, where the sum runs through all edges connecting $v_i$  and $v_j$.\\
Let $L[i]$ be the minor obtained from $L$ by removing the $i$-th row and $i$-th column. The matrix-tree theorem \cite{Tut,Sta} states that for each $i=1,\,...,\,r$ 
$$\det \left( L[i] \right)=U.$$

Obviously the theorem provides a way to derive the first Symanzik polynomial from a Laplacian matrix. As an example let us consider the graph in figure 1.

\begin{figure}[htb]
\begin{picture}(140,70)(0,0)%

\put(40,10){\epsfig{file=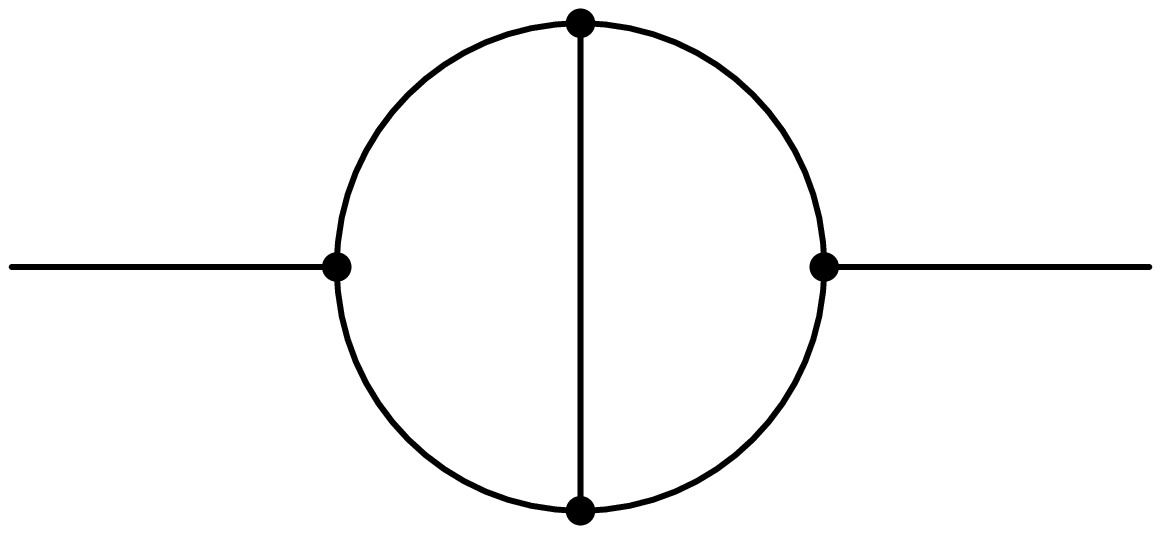,width=45mm} }

\put(78, 63) {$ {x_1} $}%
\put(122, 63) {$ {x_2} $}%
\put(123, 12) {$ {x_3} $}%
\put(77, 12) {$ {x_4} $}%
\put(108, 40) {$ {x_5} $}%

\put(100, 72) {$ {v_1} $}%
\put(133, 44) {$ {v_2} $}%
\put(100, 3) {$ {v_3} $}%
\put(67, 44) {$ {v_4} $}%

\end{picture}%

\caption{A two-loop example.}
\label{fig:twoloop}
\end{figure}

We construct the Laplacian matrix and remove the row and the column corresponding to vertex $v_4$. We obtain the minor $$L[4]=$$ $$\left(\begin{array}{ccc} x_{1}+x_{2}+x_{5} & -x_{2} & -x_{5} \\ -x_{2} & x_{2}+x_{3} & -x_{3}\\ -x_{5} & -x_{3} & x_{3}+x_{4}+x_{5}\end{array}\right).$$
The determinant of this minor is the polynomial 
\begin{eqnarray*}
x_{1}x_{2}(x_{3}+x_{4})+(x_{1}+x_{2})x_{3}x_{4}+ &  & \\
(x_{1}x_{2}+x_{1}x_{3}+x_{2}x_{4}+x_{3}x_{4})x_5 & = & U.\end{eqnarray*}
By the transformation given in the previous section, we easily obtain $\mathcal{U}$.\\
In the following we want to derive $\mathcal{F}_0$ from a Laplacian matrix. Notice that in the construction of the Laplacian matrix above we only take internal edges of the graph into account. If we want to derive the polynomial $\mathcal{F}_0$ from a Laplacian matrix, we need to appropriately incorporate the external momenta in such a matrix. This can be done by the following construction.\\
Let $G$ be a Feynman graph with $n$ internal edges $(e_1,\,...,\,e_n)$, $r$ vertices $(v_1,\,...,\,v_r)$ and $m$ external legs. At the loose ends of the legs let us attach additional 'external' vertices $v_{r+1},\,...,\,v_{r+m}$ such that the legs become internal edges of a new graph, called $\hat{G}$.  This graph has $n+m$ internal edges, $r+m$ vertices and no legs. We associate parameters $x_i$ to the first $n$ edges $e_i$ $(1\leq i\leq n)$ which correspond to the internal edges of $G$. Furthermore we assign new parameters $z_j$ to the edges $e_{n+j}$ $(1\leq j\leq m)$, corresponding to the legs of $G$. Thus each internal edge of $\hat{G}$ is labelled by one parameter in $(x_1,\,...,x_n,\,z_1,\,...,\,z_m)$. We use this set of parameters to construct the Laplacian $L(\hat{G})$  of $\hat{G}$, which is an $(r+m)\times (r+m)$ matrix whose entries are sums of (signed) variables $x_i$ and $z_i$. \\
By $L(\hat{G})[r+1,\,...,\,r+m]$ we denote the minor of $L(\hat{G})$, where the rows and columns corresponding to the 'external' vertices $v_{r+1},\,...,\,v_{r+m}$ are removed. Let us consider the polynomial $$\mathcal{W}=\det L(\hat{G})[r+1,\,...,\,r+m].$$ Minors of Laplacian matrices where more than one row and one column are removed are subject to a generalization of the matrix-tree theorem, known as all-minors matrix-tree theorem \cite{Che,Cha,Moo}. By use of this theorem it can be shown \cite{BoWe2} that the determinant $\mathcal{W}$ contains both Symanzik polynomials of $G$ in the following sense. We expand $\mathcal{W}$ in polynomials homogeneous in the variables $z_j$ as $$\mathcal{W}=\mathcal{W}^{(0)}+\mathcal{W}^{(1)}+\mathcal{W}^{(2)}+...+\mathcal{W}^{(m)},$$ where each  $\mathcal{W}^{(k)}$ is homogeneous of degree $k$ in the variables $z_j$. The first two terms are $$\mathcal{W}^{(0)}=0, \quad \mathcal{W}^{(1)}=U(x_1,\,...,\,x_n)\sum_{j=1}^m z_j.$$
The third term $\mathcal{W}^{(2)}$ is a polynomial where each term contains a product of two $z$-variables. According to the construction of $\hat{G}$, each parameter $z_i$ corresponds to an external momentum $p_i$ of $G$. Replacing each product $z_i z_j$ in $\mathcal{W}^{(2)}$ by the scalar product of the corresponding external momentum vectors  $p_i p_j$  we obtain the polynomial $F_0$: $$\mathcal{W}^{(2)}|_{z_i z_j \rightarrow p_i p_j}=F_0.$$ Applying the transformations of the previous section we arrive at $\mathcal{U}$ and $\mathcal{F}_0$ respectively.

\section{DODGSON IDENTITIES}

Let us discuss an application of the correspondence between Symanzik polynomials and determinants. A relation found by Dodgson, satisfied by arbitrary determinants, can be used for the derivation of identities between Symanzik polynomials of different graphs.\\ 
The differences between these graphs are due to the deletion and contraction of edges. Let $e$ be a regular edge of $G$, i.e.\ it is not connected to the same vertices at both ends (no tadpole) and its  removal does not increase the number of connected components of $G$ (no bridge). Then $G/e$ denotes the graph obtained from $G$ after contracting the edge $e$ while $G-e$ is the graph obtained by deleting $e$ from $G$. For any regular edge $e_k$ of $G$ the Symanzik polynomials satisfy the relations
\begin{eqnarray*}
\mathcal{U}(G) & = & \mathcal{U}(G/e_k)+x_k \mathcal{U}(G-e_k),\\
\mathcal{F}_0(G) & = & \mathcal{F}_0(G/e_k)+x_k \mathcal{F}_0(G-e_k).\end{eqnarray*}
These identities can be used to derive the Symanzik polynomials recursively.\\ 
Now let us consider an arbitrary $n\times n$ matrix $A$ and integers $i,\ j$ where $1\leq i,\,j \leq n$ and $i\neq j$. The minor obtained by removing the $i$-th row and $i$-th column is denoted $A[i]$. We furthermore write  $A[i;\,j]$ for the minor obtained by removing the $i$-th row and $j$-th column while $A[i,\,j]$ denotes the minor obtained by removing two rows and two columns numbered by $i$ and $j$ respectively. Dodgson's relation reads \cite{Dod,Zei} 
$$\det (A) \det (A[i,\,j])$$
$$=\det (A[i]) \det (A[j]) - \det (A[i;\,j]) \det (A[j;\,i]).$$ 
In order to make use of this property of determinants in the context of graph polynomials one considers a graph $G$ with two regular edges $e_a$ and $e_b$. Let $e_a$ connect the vertices $v_i$ and $v_k$ and let $e_b$ connect the vertices $v_j$ and $v_k$, i.e.\ the edges share a common vertex $v_k$. We consider the matrix $L(G-e_a-e_b)[k]$ which is the minor of the Laplacian of the graph $G-e_a-e_b$ after removing the $k$-th row and column (corresponding to vertex $v_k$). By setting $A=L(G-e_a-e_b)[k]$ in Dodgson's identity, one obtains \cite{Ste}
$$\mathcal{U}(G/e_a-e_b)\mathcal{U}(G/e_b-e_a)-$$ $$\mathcal{U}(G-e_a-e_b)\mathcal{U}(G/e_a/e_b)=\left(\frac{\Delta _1}{x_ax_b}\right)^2$$
where 
$$\Delta _1=\sum _{F\in \mathcal T_2^{(i,\,k),(j,\,k)}}\prod _{e_t\notin F}x_t.$$ 
The sum is over all spanning 2-forests of $G$ such that $v_i$ and $v_j$ are in one of the components and $v_k$ is in the other one. $\Delta _1$ is linear in each Feynman parameter. This property is used at a crucial point of Brown's algorithm \cite{Br1} for the evaluation of certain Feynman integrals to multiple zeta values. The polynomials involved in this class of identities are extensively studied in \cite{Br2,BrYe}.\\
Now let us consider the graph $\hat{G}$ again, which we constructed in order to take the external momenta into account. If we apply Dodgson's equation to an appropriate minor of the Laplacian of $\hat{G}$ we obtain an equation of the above form where on the left-hand side instead of first Symanzik polynomial the polynomial $\mathcal{W}$ appears. Expanding the polynomials $\mathcal{W}$ in powers of the $z$-variables we re-obtain at first order the above relation with the first Symanzik polynomials. The next order yields a new equation, involving the first and the second Symanzik polynomial:
$$\mathcal{U}(G/e_a-e_b)\mathcal{F}_0(G/e_b-e_a)-$$
$$\mathcal{U}(G-e_a-e_b)\mathcal{F}_0(G/e_a/e_b)+$$
$$\mathcal{F}_0(G/e_a-e_b)\mathcal{U}(G/e_b-e_a)-$$
$$\mathcal{F}_0(G-e_a-e_b)\mathcal{U}(G/e_a/e_b)$$ 
$$=2\left(\frac{\Delta _1}{x_ax_b}\right)\left(\frac{\Delta _2}{x_ax_b}\right)$$
with $\Delta_1$ as given above and where $\Delta_2$ is a polynomial generated by a certain class of spanning 3-forests of $G$. Both polynomials $\Delta_1$ and $\Delta_2$ are linear in each Feynman parameter. The definition of $\Delta_2$ and a further factorisation identity for $\mathcal{F}_0$ at a special kinematical configuration are given in \cite{BoWe2}.

\section{CYCLE MATROIDS}

It is a well-known observation that two different graphs may have the same first Symanzik polynomial. We address the question under which conditions this is the case.\\
To state this question more precisely we have to take into account that Symanzik polynomials depend on the way we distribute the Feynman parameters over the internal edges. For example let us consider the first Symanzik polynomial $U$ obtained from figure 1. If in this figure we replace $x_2$ by $x_5$ and vice versa, then we obtain a different first Symanzik polynomial, say $U'$. Such a difference, caused by re-naming variables, shall not play a role in the following. Let us say that two Symanzik polynomials are isomorphic if they can be obtained from each other after a bijection on the set of the Feynman parameters, i.e.\ a mere change of variable names as in the case of $U$ and $U'$. Obviously the polynomials obtained from the same graph are isomorphic. In which cases are the first Symanzik polynomials of different graphs isomorphic? It turns out that the answer can be obtained as a corollary of a theorem of Whitney on cycle matroids.

\begin{figure}
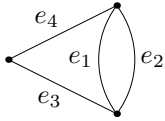

\begin{feynartspicture}(210,60)(3,1)%
\FADiagram{}

\FADiagram{}
\FAProp(-0.,7.5)(15.,0.)(0.,){/Straight}{0} \FALabel(7.3986,3.0672)[tr]{$e_3$} \FAProp(-0.,7.5)(15.,15.)(0.,){/Straight}{0} \FALabel(6.98,12.02)[br]{$e_4$} \FAProp(15.,15.)(15.,0.)(0.3333,){/Straight}{0} \FALabel(11.68,7.5)[r]{$e_1$} \FAProp(15.,15.)(15.,0.)(-0.3333,){/Straight}{0} \FALabel(18.32,7.5)[l]{$e_2$} \FAVert(-0.,7.5){0} \FAVert(15.,0.){0} \FAVert(15.,15.){0}

\FADiagram{}

\end{feynartspicture}%

\caption{A graph $G$, corresponding to a cycle matroid $M(G)=(E,\,\mathcal{I}).$ \label{fig:matroid example}}

\end{figure}

A matroid in the general sense is typically defined as a pair $(E,\,\mathcal{I})$ of a finite set $E$, the so called ground set, and a set $\mathcal{I}$ of certain subsets of $E$, called the independent sets (see e.g.\ \cite{Ox1}). In the case of a cycle matroid $M(G)$ the ground set $E$ is given by the set of edges of a graph $G$ and the independent sets are all subsets of $E$ which do not contain all the edges of a cycle of $G$. For example the cycle matroid of the graph in figure 2 is the ordered pair of 

\begin{eqnarray*}
E & = & \left\{e_1,\,e_2,\,e_3,\,e_4\right\},\\
\mathcal{I} & = & \left\{ \emptyset,\, \left\{e_1\right\},\,\left\{e_2\right\},\,\left\{e_3\right\},\,\left\{e_4\right\},\,\left\{e_1,\,e_3\right\}, \right.\\
& & \left. \left\{e_1,\,e_4\right\},\,\left\{e_2,\,e_3\right\},\,\left\{e_2,\,e_4\right\},\,\left\{e_3,\,e_4\right\}\right\}. 
\end{eqnarray*}
The sets in $\mathcal{I}$ which have the maximal number of elements are called the bases of the matroid. The independent sets $\mathcal{I}$ of a matroid are given by the bases and all possible subsets of the bases. Therefore a matroid is uniquely determined by its ground set and its bases.\\
Furthermore it can be shown that the bases of the cycle matroid of a graph $G$ are exactly the edge-sets of the spanning trees of $G$. For each base $B$ of the cycle matroid $M(G)$ there is a spanning tree of $G$ whose edges are the members of $B$. Therefore we can write the first Symanzik polynomial of $G$ as 
$$U=\sum_{B_j\in \mathcal{B}} \; \prod_{e_i\in B_j}x_i,$$
where the sum is over the set of bases $\mathcal{B}$ of the cycle matroid of $G$.\\
As a consequence of the correspondence between the spanning trees and the bases of the cycle matroid we can re-formulate our question: In which cases are the cycle matroids of different graphs isomorphic? For connected graphs these are exactly the cases when the first Symanzik polynomials are isomorphic.\\
The answer is given by Whitney's 2-isomorphism theorem \cite{Whi,Ox2,Tru}: Let $G$ and $H$ be two graphs without isolated vertices. Their cycle matroids are isomorphic if and only if $G$ is obtained from $H$ after a sequence of the three transformations of (1) vertex identification, (2) vertex cleaving and (3) twisting.\\ 
Let us briefly discuss these transformations. Consider a graph $G$ with vertices $u$ and $v$ belonging to two different components. Vertex identification gives a new graph $G'$ obtained by identification of $u$ and  $v$ as a new vertex in $G'$. The reverse process where we obtain $G$ from $G'$ is vertex cleaving. An example for the third transformation is shown in figure 3 where $G'$ is obtained from $G$ by twisting about $u$ and $v$ and vice versa. $G$ is obtained from disjoint graphs $G_1$ and $G_2$ by identifying $u_1$ with $u_2$ and $v_1$ with $v_2$. $G'$ instead is obtained from the same two graphs by identifying $u_1$ with $v_2$ and $v_1$ with $u_2$. By sequences of these three transformations the connected graphs with isomorphic first Symanzik polynomials are obtained from each other.

\begin{figure}
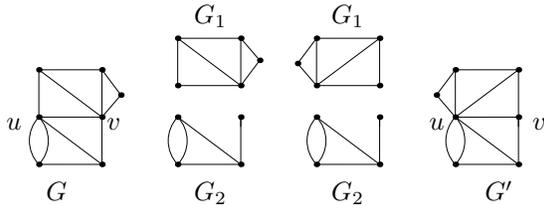

\begin{feynartspicture}(210,75)(4,1)%

\FADiagram{}
\FAProp(5.,15.)(15.,15.)(0.,){/Straight}{0}  \FAProp(5.,15.)(5.,7.5)(0.,){/Straight}{0}  \FAProp(5.,7.5)(15.,7.5)(0.,){/Straight}{0}  \FAProp(15.,15.)(15.,7.5)(0.,){/Straight}{0}  \FAProp(5.,7.5)(15.,0.)(0.,){/Straight}{0}  \FAProp(5.,-0.)(15.,0.)(0.,){/Straight}{0}  \FAProp(5.,-0.)(5.,7.5)(-0.4,){/Straight}{0} \FAProp(15.,7.5)(15.,0.)(0.,){/Straight}{0}  \FAProp(15.,15.)(18.,11.)(0.,){/Straight}{0}  \FAProp(15.,7.5)(18.,11.)(0.,){/Straight}{0}  \FAProp(5.,7.5)(5.,-0.)(-0.4,){/Straight}{0}  \FAProp(5.,15.)(15.,7.5)(0.,){/Straight}{0}  \FAVert(5.,15.){0} \FAVert(15.,15.){0} \FAVert(5.,7.5){0} \FAVert(15.,7.5){0} \FAVert(5.,-0.){0} \FAVert(15.,0.){0} \FAVert(18.,11.){0}

\FADiagram{}

\FAProp(5.,7.5)(15.,0.)(0.,){/Straight}{0}  \FAProp(5.,-0.)(15.,0.)(0.,){/Straight}{0}  \FAProp(5.,-0.)(5.,7.5)(-0.4,){/Straight}{0}  \FAProp(15.,7.5)(15.,0.)(0.,){/Straight}{0}  \FAProp(15.,20.)(18.,16.5)(0.,){/Straight}{0}  \FAProp(5.,7.5)(5.,-0.)(-0.4,){/Straight}{0}  \FAProp(5.,20.)(15.,12.5)(0.,){/Straight}{0}  \FAProp(5.,20.)(15.,20.)(0.,){/Straight}{0}  \FAProp(18.,16.5)(15.,12.5)(0.,){/Straight}{0}  \FAProp(15.,20.)(15.,12.5)(0.,){/Straight}{0}  \FAProp(15.,12.5)(5.,12.5)(0.,){/Straight}{0}  \FAProp(5.,12.5)(5.,20.)(0.,){/Straight}{0}  \FAProp(15.,7.5)(15.,6.)(0.,){/Straight}{0}  \FAVert(15.,20.){0} \FAVert(5.,7.5){0} \FAVert(15.,7.5){0} \FAVert(5.,-0.){0} \FAVert(15.,0.){0} \FAVert(5.,20.){0} \FAVert(15.,12.5){0} \FAVert(18.,16.5){0} \FAVert(5.,12.5){0}

\FADiagram{}

\FAProp(5.,7.5)(15.,0.)(0.,){/Straight}{0}  \FAProp(5.,-0.)(15.,0.)(0.,){/Straight}{0}  \FAProp(5.,-0.)(5.,7.5)(-0.4,){/Straight}{0}  \FAProp(15.,7.5)(15.,0.)(0.,){/Straight}{0}  \FAProp(5.,7.5)(5.,-0.)(-0.4,){/Straight}{0}  \FAProp(5.,20.)(15.,20.)(0.,){/Straight}{0}  \FAProp(15.,20.)(15.,12.5)(0.,){/Straight}{0}  \FAProp(15.,12.5)(5.,12.5)(0.,){/Straight}{0}  \FAProp(5.,12.5)(5.,20.)(0.,){/Straight}{0}  \FAProp(15.,7.5)(15.,6.)(0.,){/Straight}{0}  \FAProp(5.,12.5)(15.,20.)(0.,){/Straight}{0}  \FAProp(5.,20.)(2.,16.)(0.,){/Straight}{0}  \FAProp(5.,12.5)(2.,16.)(0.,){/Straight}{0}  \FAVert(15.,20.){0} \FAVert(5.,7.5){0} \FAVert(15.,7.5){0} \FAVert(5.,-0.){0} \FAVert(15.,0.){0} \FAVert(5.,20.){0} \FAVert(15.,12.5){0} \FAVert(5.,12.5){0} \FAVert(2.,16.){0}

\FADiagram{}

\FAProp(5.,7.5)(15.,0.)(0.,){/Straight}{0}  \FAProp(5.,-0.)(15.,0.)(0.,){/Straight}{0}  \FAProp(5.,-0.)(5.,7.5)(-0.4,){/Straight}{0}  \FAProp(15.,7.5)(15.,0.)(0.,){/Straight}{0}  \FAProp(5.,7.5)(5.,-0.)(-0.4,){/Straight}{0}  \FAProp(5.,15.)(15.,15.)(0.,){/Straight}{0}  \FAProp(15.,15.)(15.,7.5)(0.,){/Straight}{0}  \FAProp(15.,7.5)(5.,7.5)(0.,){/Straight}{0}  \FAProp(5.,7.5)(5.,15.)(0.,){/Straight}{0}  \FAProp(15.,7.5)(15.,6.)(0.,){/Straight}{0}  \FAProp(5.,7.5)(15.,15.)(0.,){/Straight}{0}  \FAProp(5.,15.)(2.,11.)(0.,){/Straight}{0}  \FAProp(5.,7.5)(2.,11.)(0.,){/Straight}{0}  \FAVert(15.,15.){0} \FAVert(5.,7.5){0} \FAVert(15.,7.5){0} \FAVert(5.,-0.){0} \FAVert(15.,0.){0} \FAVert(5.,15.){0} \FAVert(2.,11.){0}

\put(2, 27) {$ {u} $}%
\put(40, 27) {$ {v} $}%

\put(162, 27) {$ {u} $}%
\put(201, 27) {$ {v} $}%

\put(73, 67) {$ {G_1} $}%
\put(125, 67) {$ {G_1} $}%
\put(73, 0) {$ {G_2} $}%
\put(125, 0) {$ {G_2} $}%
\put(17, 0) {$ {G} $}%
\put(183, 0) {$ {G'} $}%

\end{feynartspicture}%

\caption{Twisting about $u$ and $v$.\label{fig:twisting}}

\end{figure}

\subsection*{Acknowledgements} I want to thank Stefan Weinzierl for his encouragement and the joint work \cite{BoWe2} which this talk is based on. I am thankful for communication with Francis Brown, Eric Patterson and Karen Yeats. This work was supported by Deutsche Forschungsgemeinschaft SFB/TR9.

\end{document}